\documentclass[iop,natbib]{emulateapj}

\usepackage[]{natbib}
\usepackage{epstopdf}
\newcommand{\ubo}{($U$-$B$)$_0$}
\newcommand{\av}{$A_V$}
\newcommand{\actaa}{Acta Astronomica}

\newcommand{\logl}{log($L$/$L_{\odot}$)}

\newcommand{\teff}{$T_{\rm eff}$}

\newcommand{\mv}{$M_V$}
\newcommand{\bvo}{$(B - V)_0$}

\newcommand{\ebv}{$E(B-V)$}
\newcommand{\mjup}{$M_{\rm Jup}$}
\newcommand{\msun}{$M_{\odot}$}
\newcommand{\rsun}{$R_{\odot}$}

\newcommand{\logg}{log($g$)}

\newcommand{\evi}{$E$($V - I_C$)}

\slugcomment{submitted to Astrophysical Journal}
\shorttitle{}
\shortauthors{}
\begin{document}

\title{Modeling Transiting Circumstellar Disks: Characterizing the
  Newly Discovered Eclipsing Disk System OGLE LMC-ECL-11893}

\author{ 
Erin L. Scott\altaffilmark{1},
Eric E. Mamajek\altaffilmark{1}, 
Mark J. Pecaut\altaffilmark{1,2}, 
Alice C. Quillen\altaffilmark{1}, 
Fred Moolekamp\altaffilmark{1}, and
Cameron P. M. Bell\altaffilmark{1}}

\altaffiltext{1}{Department of Physics and Astronomy, 
University of Rochester, Rochester, NY 14627-0171, USA}
\altaffiltext{2}{Physics Department, Rockhurst University, 
1100 Rockhurst Road, Kansas City, MO 64110, USA}
\email{elscott@pas.rochester.edu}

\begin{abstract}

We investigate the nature of the unusual eclipsing star OGLE
LMC-ECL-11893 (OGLE J05172127-6900558) in the Large Magellanic Cloud
recently reported by Dong {\it et al.} The eclipse period for this
star is 468 days, and the eclipses exhibit a minimum of $\sim$1.4 mag,
preceded by a plateau of $\sim$0.8 mag. Spectra and optical/IR
photometry are consistent with the eclipsed star being a lightly
reddened B9III star of inferred age $\sim$150 Myr and mass $\sim$4
$M_{\odot}$. The disk appears to have an outer radius of $\sim$0.2 AU
with predicted temperatures of $\sim$1100-1400\,K. We model the
eclipses as being due to either a transiting geometrically thin dust
disk or gaseous accretion disk around a secondary object; the debris
disk produces a better fit.  We speculate on the origin of such a
dense circumstellar dust disk structure orbiting a relatively old
low-mass companion, and on the similarities of this system to the
previously discovered EE Cep.
\end{abstract}

\section{Introduction}

In the discovery paper for the eclipsing circumsecondary disk system
orbiting 1SWASP J140747.93-394542.6, \citet{Mamajek12} explored the
possibility of finding other such systems, calculating the probability 
to be nonnegligable given continuous monitoring of a large number of
young stars. Examples of eclipsing disks are notably rare, including
famous examples of $\epsilon$ Aur, EE Cep, and OGLE-LMC-ECL-17782
\citep[e.g.,][]{Kloppenborg10, Galan12, Graczyk11}, and recent surveys
have yielded few, if any, candidates \citep[e.g.,][]{Meng14}. Modeling
of eclipsing disks can provide constraints on the geometry and
structure of circumstellar disks--details which are difficult to
constrain from spectral energy distributions alone.


OGLE 11893 (OGLE J05172127-6900558 = LMC SC8 354550; hereafter ``OGLE
11893'')\footnote{The Optical Gravitational Lensing experiment,
  http://ogle.astrouw.edu.pl/ \citep{Udalski00}.} was first cataloged
as a LMC member by \citet{Udalski00}, flagged as a variable by
\citet{Zebrun01}, and classified as an eclipsing variable by
\citet{Graczyk11}. \citet{Dong14} first reported OGLE 11893 to 
be a particularly unusual eclipsing binary with a long period, 
deep eclipse, and asymmetric shape. \citet{Dong14} proposed that 
the eclipses could be due to a transiting circumstellar disk. The 
properties of OGLE 11893 are compiled in Table \ref{tab:star}. The 
eclipses are a repeating phenomenon, with a period of 468 days and 
depth of $\sim$1.4 mag \citep{Graczyk11, Dong14}. The phase-folded 
light curve reveals an eclipse structure that is highly consistent 
on the timescale over which the star has been observed 
\citep[$\sim$17 yr, $\sim$13 orbital periods;][]{Dong14}, 
indicating that the disk is not precessing at a detectable rate. 
\citet{Dong14} reported preliminary observations from the 
H$\alpha$ line profile that the star appears to be a Ae/Be star 
exhibiting H$\alpha$ emission.

In this contribution, we further characterize OGLE 11893 and 
model its eclipses using circumstellar disk models. In Section 2 
we discuss observations for the star. In Section 3 we analyze the 
properties of OGLE 11893 and its light curve, and model the 
eclipses as due to various flavors of disks.  In Section 4 we 
summarize our findings regarding the eclipsing disk.

\section{Data \label{sec:data}}

\subsection{Photometry \label{sec:phot}}

In order to model the light curve, we used data from the photometric
observations of the OGLE-II and OGLE-III surveys, spanning 12 yr 
in total. The OGLE-II survey took place from 1996 December to 2000 
November, using standard $V$ and $I$ photometric bands and generating
300-500 frames per field in $I$ \citep{Wyrzykowski09}. OGLE-III took 
place from 2001 July to 2009 May and produced a total of 866 photometric 
points for OGLE 11893 in $I$ band and 61 points in $V$ over the course 
of 8 yr \citep{Graczyk11}. Further details on the star can be found 
in \citet{Dong14} and in Table \ref{tab:phot}.

\subsection{Spectroscopy \label{sec:spec}}

\begin{figure}
\begin{center}
\includegraphics[scale=0.4]{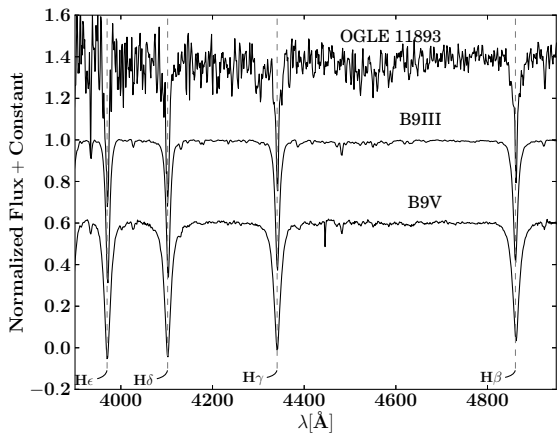}
\caption{\label{fig:bluespectrum} 
  Comparison of the IMACS spectrum of
  OGLE~11893 to that of spectral standards HD~141774 (B9 V) and
  $\gamma$~Lyr (B9 III).  Here the B9 V standard is HD~141774
  \citep{Garrison79} and the B9 III standard is $\gamma$~Lyr
  \citep{Johnson53}.}
\end{center}
\end{figure}

We analyze the blue optical spectrum of OGLE-LMC-ECL-11893 (Figure 
\ref{fig:bluespectrum}) presented in \citet{Dong14}, taken by J. 
Prieto on UT 2011 December 23 with the IMACS \citep[Inamori-Magellan 
Areal Camera \& Spectrograph;][]{Dressler06} instrument on the Baade 
6.5 m telescope at Las Campanas Observatory. The blue spectrum was 
taken with the long 0''.7 slit and 300 L mm$^{-1}$ grism, and yielded 
resolution $R$ $\sim$ 1800 between $\lambda\lambda$ 3500-6574 \AA. 
The S/N of the IMACS blue spectrum is 
$\sim$22 ($\sim$5000\AA). Further details are described in Section 2 
of \citet{Dong14}.

\begin{deluxetable*}{lcll}[htb!]
\tabletypesize{\scriptsize}
\setlength{\tabcolsep}{0.03in}
\tablewidth{0pt}
\tablecaption{Data for OGLE LMC-ECL-11893\label{tab:star}}
\tablehead{
{(1)}       & {(2)}   & {(3)}   & {(4)}\\
{Parameter} & {Value} & {Units} & {Ref.}}
\startdata
$\alpha$(ICRS) & 05:17:21.27          & h:m:s & \citet{Udalski00}\\
$\delta$(ICRS) & -69:00:55.8          & d:m:s & \citet{Udalski00}\\
Spec. Type     & B9III:               & ... & This paper\\
$U$            & 17.854\,$\pm$\,0.060 & mag & \citet{Zaritsky04}\\
$B$            & 17.954\,$\pm$\,0.021 & mag & \citet{Udalski00}\\
$V$            & 17.753\,$\pm$\,0.01   & mag & \citet{Graczyk11}\\
$I_c$          & 17.556\,$\pm$\,0.01   & mag & \citet{Graczyk11}\\
$J$            & 17.61\,$\pm$\,0.04   & mag & \citet{Kato07}\\
$H$            & 17.39\,$\pm$\,0.10   & mag & \citet{Kato07}\\
$K_s$          & 17.43\,$\pm$\,0.23   & mag & \citet{Kato07}\\
$[$3.6$]$      & 16.932\,$\pm$\,0.107 & mag & \citet{Meixner06}\\
$[$4.5$]$      & 16.955\,$\pm$\,0.104 & mag & \citet{Meixner06}\\
Period         & 468.045440           & day & \citet{Graczyk11}\\
Epoch(min.,HJD) & 2454011.5436     & day & \citet{Graczyk11}\\
Dist. Mod.(LMC) & 18.48\,$\pm$\,0.05  & mag & \citet{Walker12}\\
$R_V$(LMC)      & 3.41\,$\pm$\,0.06   & ... & \citet{Gordon03}\\
\teff\,         & 12000\,$\pm$\,300   & K   & This paper\\
$E(B-V)$          & 0.28\,$\pm$\,0.02   & mag & This paper\\
$A_V$           & 0.91\,$\pm$\,0.06   & mag & This paper\\
$M_V$           & -1.64\,$\pm$\,0.08  & mag & This paper\\
$M_{\rm bol}$     & -2.32\,$\pm$\,0.10  & mag & This paper\\
\logl\,        & 2.83\,$\pm$\,0.04   & dex & This paper\\
Radius         & 6.03\,$\pm$\,0.42   & \rsun & This paper\\
Mass           & $\sim$4.1           & \msun & This paper\\
Age            & $\sim$150           & Myr   & This paper
\enddata
\label{tab:phot}
\tablecomments{We assume $\pm$0.01 mag uncertainty in the mean
  out-of-eclipse $V$ and $I$ magnitudes from \citet{Graczyk11};
  [3.6] and [4.5] {\it Spitzer} IRAC photometry is from the SAGE survey 
  \citep{Meixner06} catalog accessible via Vizier at:
  http://vizier.cfa.harvard.edu/viz-bin/Cat?II/305.}
\end{deluxetable*}

\section{Analysis \label{sec:anal}}

\subsection{Light Curve \label{sec:lightcurve}}

\begin{figure*}[htp]
\begin{center}
\includegraphics[scale=0.4]{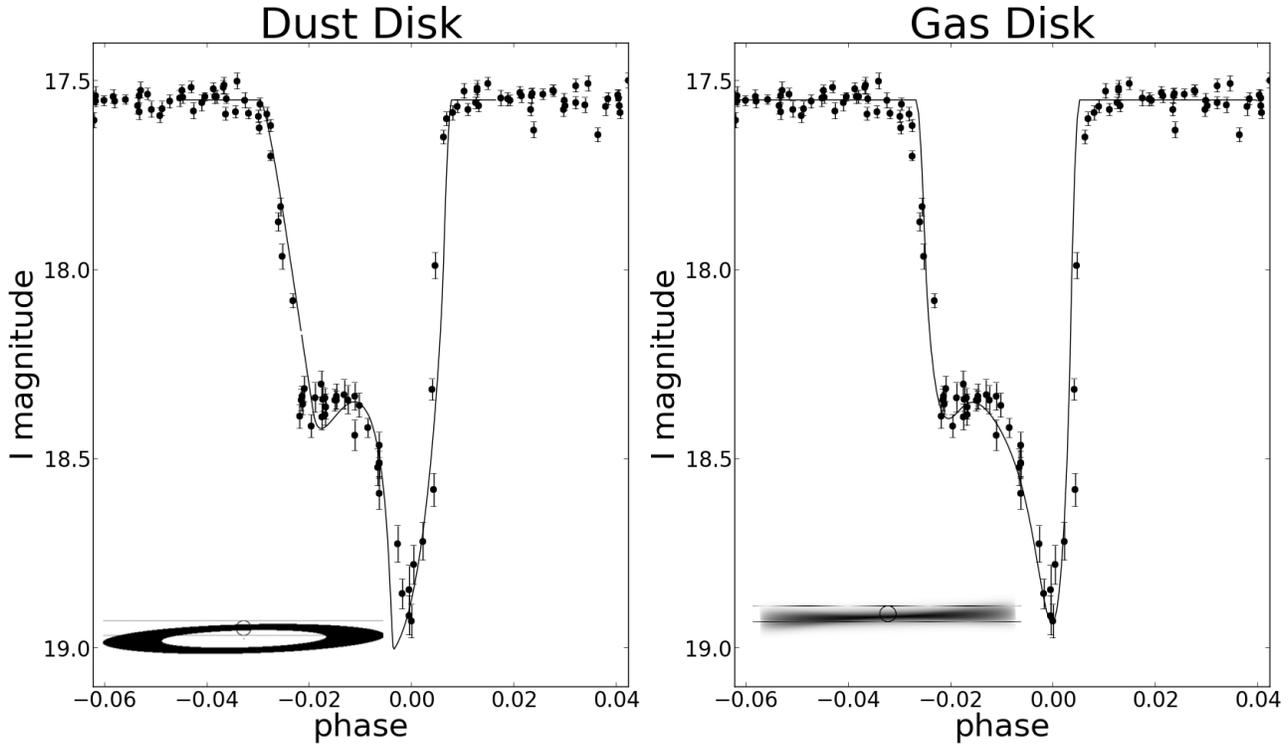}
\caption{Phase-folded light curve for OGLE~11893 with best-fit 
synthetic light curves for ({\it left}) an infinitely thin dusty 
debris disk, and ({\it right}) flared gas-rich disk. Parameters
of best fit synthetic light curves are compiled in Table \ref{tab:fits}.
\label{fig:model}}
\end{center}
\end{figure*}

Figure \ref{fig:model} shows the phase-folded light curve of OGLE~11893.
The light curve shows a dimming of $\sim$0.8 mag over $\sim$8
days, which deepens to $\sim$1.4 mag before returning to full
brightness.  The total duration of the eclipse is $\sim$16 days. As we
will show, this pattern is consistent with a disk that is nearly
edge-on to the observer's line of sight, and which has a cleared inner
region.

\subsection{Spectral Classification \label{sec:spt}}

As shown in Figure \ref{fig:bluespectrum}, through comparison of 
the IMACS spectrum of OGLE~11893 to a grid of Galactic MK standards 
\citep{Morgan73} observed at a resolution of 4.3\text{\AA} with 
the RC spectrograph on the SMARTS 1.5 m telescope, we find that 
the Balmer lines are consistent with that of a B9 giant star. The 
luminosity class, however, is still somewhat uncertain due to the 
low S/N of the spectrum.

\subsection{Reddening \label{sec:red}}

We estimate the reddening by comparing to examples of unreddened
Galactic late B-type giants.  \citet{Garrison94} classified 10 
stars as B9III or variant. Of those stars, the revised {\it Hipparcos} 
parallaxes of \citet{vanLeeuwen07} only place two within 100 pc 
(and hence negligibly reddened): HD 145389 and HD 129174. Using 
the \citet{vanLeeuwen07} parallaxes, and $UBV$ photometry from 
\citet{Mermilliod97}, we estimate their mean properties as \bvo\, 
$\simeq$ -0.09\,$\pm$\,0.03, \ubo\, $\simeq$ -0.329\,$\pm$\,0.10, 
\mv\, $\simeq$ 0.20\,$\pm$\,0.08. Based solely on comparison of 
OGLE~11893's spectral type and colors in Table \ref{tab:phot} to 
those of these well-studied nearby B9~III templates, we would expect 
$E(B-V) \simeq$ 0.29\,$\pm$\,0.05, $A_V$ $\simeq$0.94\,$\pm$\,0.17 
\citep[assuming $E(B-V)-A_V$ relation of][]{Olson75}, and distance 
modulus $(m-M)_o$ $\simeq$ 16.6. Hence if it were comparable to 
Galactic B9~III stars, we would predict a distance of $\sim$21 kpc 
\citep[well short of the LMC's distance of $\sim$50 kpc;][]{Walker12}. 
The reddening is somewhat larger than predicted for the star's 
position in the LMC.

We can also estimate the intrinsic color of the B-type star using 
the $Q$-method with the observed $UBV$ photometry. We assume that the 
star suffers visual reddening similar to that of the Galactic average 
value \citep[$E(U-B)$/$E(B-V)$ = 0.72; ][]{Johnson53}, which is statistically
consistent with values derived for hot stars in the LMC not near 
the 30 Dor complex \citep{Walker68, Fitzpatrick86, Gordon03}. Using 
the revised calibration of $Q$ versus spectral type for Be stars from
\citet{Halbedel93}, and the observed colors (see Table \ref{tab:phot};
\ub\, = -0.10\,$\pm$\,0.06, \bv\, $\simeq$ 0.20\,$\pm$\,0.04), we
estimate the star to have $Q$ = -0.245 and a photometric spectral type
of B8.3e. Unfortunately \cite{Halbedel93} does not provide the
intrinsic colors for the Be stars. However, if we adopt the intrinsic
color sequence for B stars from \citet{Pecaut13}, the dereddened
photometry corresponds to \bvo\, = -0.095, \ubo\,= -0.314, $E(B-V)$ =
0.296. According to the updated spectral type-effective temperature
relations of \citet{Pecaut13}, for a Galactic main sequence star, this
would correspond to a $\sim$B8 star with \teff\, $\simeq$ 12100 K.

The foreground Galactic reddening to the LMC varies from $E(B-V)
\simeq$ 0.04-0.09 \citep{Bessell91}. Typical reddenings for hot 
stars in the LMC are $E(B-V)\simeq$ 0.15-0.31 mag \citep{Gordon03}, 
consistent with what we see for OGLE 11893.  However this is at 
odds with the extinction inferred from analyses of LMC RR Lyraes 
and red clump giants \citep{Pejcha09,Haschke11}, which yield 
$E(V-I_c$) = 0.153 and 0.070, respectively, in the vicinity of 
OGLE 11893 \citep[corresponding to $A_V$ $\simeq$ 0.38 and 0.17, 
adopting \av/\evi\, = 2.49;][]{Stanek96}. So while OGLE 11893 
appears to have a typical reddening compared to hot stars studied 
in the LMC, it appears to be anomalously redder compared to that 
predicted from reddening maps of particular classes of star (e.g. 
RR Lyraes).

Using the SED fitting routine described in \citet{Pecaut13}, we 
fit the $UBVIJHK_s$ photometry of OGLE 11893 through comparison 
to a grid of colors of Galactic B-type dwarfs of varying \ebv\, 
(assuming a Galactic reddening curve).  The best fit SED was for 
a star with \bvo\, = -0.093, \ebv\, = 0.260 mag, \av\, = 0.84 
mag, and \teff\,$\simeq$ 11800\,K. These values are similar to 
what was previously derived using the $Q$-method.

\subsection{HR Diagram Position}

Based on the spectrum and photometry, we adopt for this star \teff\,
$\simeq$ 12000\,$\pm$\,300 K, with reddening $E(B-V)$ =
0.28\,$\pm$\,0.02 mag, extinction $A_V$ = 0.91\,$\pm$\,0.06 mag.
Adopting the mean LMC distance modulus from \citet{Walker12} of
18.48\,$\pm$\,0.04 mag ($\varpi$ = 0.0201\,$\pm$\,0.0004 mas), and a
\teff-appropriate bolometric correction from \citet{Code76} of BC$_V$
= -0.68\,$\pm$\,0.06 mag, we derive \mv\, = -1.64\,$\pm$\,0.08,
$M_{\rm bol}$ = -2.32\,$\pm$\,0.10, \logl\, = 2.83\,$\pm$\,0.04, and
radius 6.03\,$\pm$\,0.42 \rsun.

\subsection{Age and Primary Mass}

In order to estimate an age and mass based on HR diagram position, we use 
the PARAM 1.2 program,\footnote{http://stev.oapd.inaf.it/cgi-bin/param} 
which uses the \citet{Bressan12} evolutionary tracks. Adopting \teff\, 
= 12000\,$\pm$\,300 K, [Fe/H] = -0.3\,$\pm$\,0.05 dex \citep[typical for 
LMC; ][]{Luck98}, $V_o$ = 16.84\,$\pm$\,0.06 mag, and LMC parallax from 
\citet{Walker12}, and assuming Bayesian priors of a \citet{Chabrier01}
lognormal IMF and constant star-formation rate, PARAM 1.2 estimates the 
following a posteriori stellar parameters: age 154\,$\pm$\,8 Myr, 
mass = 4.09\,$\pm$\,0.14 \msun, \logg\, = 3.50\,$\pm$\,0.02 dex, radius 
5.76\,$\pm$\,0.14 \rsun. Using the piecewise linear fit from \citet{Ekstrom12}, 
including rotation, we derive a main sequence lifetime of $\sim$185 Myr, 
hence OGLE 11893 is closer to the end of its lifetime than the beginning.

Given a period of 468 days and a primary mass of 4.09 \msun, if we
assume that the mass of the secondary is negligible compared to the
primary, the orbital separation of the secondary is approximately 1.7
AU. Using the eclipse duration and the orbital period, we infer that
the radius of the disk is $\sim$0.20 AU. The disk radius should be
smaller than the Hill radius of the secondary, placing a lower limit
on the mass of the secondary of $M_2 \simeq 3M_1(r_{\rm disk}/a)^3 >$ 
0.014 \msun\, $\simeq$ 15 \mjup. Hence the disked secondary is almost 
certainly a star or brown dwarf rather than a giant planet. Given the 
high luminosity of the primary, we find that the equilibrium disk 
temperature due to the primary alone is probably $T_{\rm eq}$ $\simeq$1250\,K, 
not far below the sublimation temperatures of silicate dust.

\subsection{Eclipse Modeling}

The disk structure was analyzed using a code\footnote{Written 
by Fred Moolekamp, maintained and updated by Erin Scott (UR).} 
that uses a model to generate synthetic light curves for varying 
parameters. The object-oriented model is made up of a star, a 
secondary object, and a disk. The star, which is held at a fixed 
mass, is modeled as a spherical grid of points, each of which 
contains a fraction of the star's total flux, using the limb 
darkening values from \citet{vanHamme93}. The secondary is modeled 
as a solid sphere. Our calculations show that the secondary is 
between 0.15 and 0.38 \msun\ (see Equations (\ref{eqn:secmass}) 
and (\ref{sec:gas})). To calculate the secondary radius and 
temperature, we adopted an isochrone from \citet{Bressan12}, which 
covers mass ranges from 0.1 to 3.8 \msun, adopting LMC composition 
$Y$ = 0.2572, $Z$ = 0.0047 for age 200 Myr.

These objects are put together into an orbital system, the 
parameters of which are the period and the inclination with 
respect to the observer's line of sight. To avoid degeneracy 
in the best-fit solution, we assume a perfectly circular orbit.
The program then uses vector addition to determine whether any 
point on the star is eclipsed by any part of the secondary or 
disk as seen by the observer. The observed magnitude is then 
calculated and recorded.

To find a model that best fits the observed data, the program 
utilizes a simplex routine from \citet{NumericalRecipes} to 
minimize $\chi^2$. Due to a systematic variation in the error, 
which was consistently larger for dimmer points in the light 
curve, resulting in overfitting in the ingress and egress regions 
of the eclipse and underfitting in the central regions, all 
errors were set equal to the median value of 0.02 mag.

The free parameters are the orientation of the disk ($\theta_x$ and
$\theta_y$), the dimensions of the disk ($R_{\rm in}$ and $R_{\rm out}$), 
and the inclination of the orbit. If the model is of a flared gas 
disk rather than a flat dust disk, the inner radius is automatically 
set to the dust sublimation radius, but additional parameters are also
allowed to vary: the density power law 1 $\leq \alpha \leq 3$ (where
$\rho \propto r^{-\alpha}$), the secondary mass 0.15 \msun $\leq$
$M_2$ $\leq$ 0.38 \msun whose bounds were determined by calculating
the Hill radius (see Equations (\ref{eqn:hill}) and (\ref{eqn:secmass})),
and the disk mass $M_{\rm disk}$, which was varied on a logarithmic 
scale of $-2 \leq \gamma \leq-11$, where $\gamma =$
log$\left(M_{\rm disk}/M_2\right)$. At the present time we are unable 
to constrain the mass opacity
  of the disk matter, so we adopt a realistic opacity from
  \citet{Miyake93} of $\kappa\,=\,0.1$ m$^2$\,kg$^{-1}$ which would
  correspond to the $V$-band mass opacity of a population of compact
  dust particles with composition representative of interstellar dust,
  grain sizes between 10$^{-8}$ and 10$^{-1}$\,m following a
  size$^{-3.5}$ power-law distribution.  Examination of the multi-band
  light curve from \citet{Dong14} shows that the transiting dust
  appears to be consistent with $A_B$/$A_V$ $\simeq$ 1.2, which is
  close to that expected for a $R_V$\,$\simeq$\,5 interstellar dust
  extinction curve \cite{Mathis90}. The transiting dust is clearly not
  gray, and likely is reflecting a power-law distribution of dust
  grains down to submicron size, not too unlike an interstellar dust
  grains population.

\subsubsection{Flat Dust Disk}

The eclipse was first fit with the simplest model, an infinitely thin
annulus. The best fit for this flat dust disk was found to be an inner 
radius of $R_{\rm in}$ = 26.2 $R_\odot$, an outer radius of $R_{\rm out}$ = 
45.8 $R_\odot$, a disk tilt of $\theta_x = 2.0^{\circ}$, a disk obliquity 
of $\theta_y = 7.0^{\circ}$ from edge-on, an orbital inclination of $i = 
89.38^{\circ}$, and a flat optical depth of $\tau_0=1.8$. This simple 
disk produced a model light curve that was a reasonable visual fit to 
the data, with $\chi^2 / \nu$ of $1186 / 113 = 10.5$.

A slightly more complex model for the debris disk allowed the optical 
depth to vary according to a power law, which produced a best fit model 
with $R_{\rm in}$ = 26.1 $R_\odot$, an outer radius of $R_{\rm out}$ = 44.8 
$R_\odot$, a disk tilt of $\theta_x = 1.5^\circ$, a disk obliquity of 
$\theta_y = 5.9^\circ$ from edge-on, an orbital inclination of $i = 
89.46^\circ$, and an optical depth of $\tau_0 = 7.9$ at the inner edge 
of the disk which varied according to a power law of $p = 0.6$. This 
model had $\chi^2 / \nu = 1073.8 / 112 = 9.6$, and was a better overall 
fit to the data.

The infinitely thin debris disk model captures the overall light curve
structure relatively well, but has some trouble matching the data in
the central regions of the eclipse (see Figure \ref{fig:model}).

\subsubsection{Flared Gas and Dust Disk \label{sec:gas}}

Since the lack of detailed substructure indicates the possibility that
the eclipses are due to a disk with non-negligible scale height, we
next attempted to fit the light curve with a flared gas and dust disk
model. Assuming vertical hydrostatic equilibrium, the scale height 
of such a disk is given by \citet{Shakura73}:

\begin{equation}
\label{eqn:SH}
h\left(r\right) = \frac{c_s}{\Omega}
\end{equation}

\noindent where $c_s$ is the sound speed

\begin{equation}
\label{eqn:sound}
c_s = \sqrt{\frac{k_bT\left(r\right)}{\mu m_H}}
\end{equation}

\noindent and, assuming circular Keplerian orbits, $\Omega$ is the 
rotation rate

\begin{equation}
\label{eqn:rotation}
\Omega = \sqrt{\frac{GM_2}{r^3}},
\end{equation}

\noindent which gives a scale height of

\begin{equation}
\label{eqn:SHfinal}
h\left(r\right) = \sqrt{\frac{k_br^3T\left(r\right)}{\mu m_HGM_2}}.
\end{equation}

The temperature profile can be simplified for modeling purposes.
Given the size of the disk, the outer disk is almost certainly
dominated by heating from the primary star, as we discuss in Section 
3.5, putting it at a temperature of $T_{\rm out}$ $\simeq$ 1100\,K. 
The inner edge of the disk is likely at the silicate dust sublimation
temperature $T_{\rm in}$ $\simeq$ 1400\,K, as commonly seen among T
Tauri disks \citep{Muzerolle03}. Hence the radial variation in
temperature throughout the disk is likely less than $\sim$25\%. 
Since $h \propto T^{1/2}$, the scale height $h$ likely varies by 
only order $\sim$15\% between the inner and outer radii. To simplify 
our calculations, we adopt $T$ = 1250\,K, which should translate 
to accuracy of scale heights $\sim$10\%\,--15\%\, over the whole disk.

For a roughly constant temperature, the density structure 
of a flared accretion disk can be modeled by a Gaussian curve 
\citep{Shakura73}:


\begin{equation}
\label{eqn:density}
\rho\left(r,z\right) = \rho\left(r,0\right)e^{z^2/2h(r)^2}
\end{equation}

\noindent{where $1\leq\alpha\leq3$. The midplane density $\rho_0$ 
is set so as to give the correct disk mass\footnote{The density 
calculation only accounts for the small dust grains that are the 
primary source of opacity. Gas and larger grains that contribute 
most of the mass, but are not a significant source of opacity, 
are not included in the calculation, resulting in a model disk 
mass that is orders of magnitude lower than the actual mass.} 
for a given value of $\alpha$. Assuming that the temperature of 
the disk is constant, the density equation can be integrated:}

\begin{equation}
M_{\mathrm{disk}} = \int_{R_{\mathrm{in}}}^{R_{\mathrm{out}}}\int_{-\infty}^\infty\int_0^{2\pi}\rho_0e^{z^2/2h(r)^2}\left(\frac{r}{R_2}\right)^{-\alpha}r\mathrm{d}r\mathrm{d}\phi\mathrm{d}z
\end{equation}

\begin{equation}
M_{\mathrm{disk}} = 2\pi\rho_0\int_{R_{\mathrm{in}}}^{R_{\mathrm{out}}}r\left(\frac{r}{R_2}\right)^{-\alpha}\mathrm{d}r\int_{-\infty}^\infty e^{z^2/2h(r)^2}\mathrm{d}z
\end{equation}

\begin{equation}
M_{\mathrm{disk}} = (2\pi)^{\frac{3}{2}}\rho_0R_2^\alpha\int_{R_{\mathrm{in}}}^{R_{\mathrm{out}}}r^{1-\alpha}h(r)\mathrm{d}r
\end{equation}

\begin{equation}
M_{\mathrm{disk}} = \frac{2(2\pi)^{3/2}}{7-2\alpha}\rho_0R_2^\alpha\left[R_{\mathrm{out}}^{2-\alpha}h\left(R_{\mathrm{out}}\right) - R_{\mathrm{in}}^{2-\alpha}h\left(R_{\mathrm{in}}\right)\right],
\end{equation}

\noindent which can then be solved for $\rho_0$.

In this model, the secondary mass becomes important, as it determines
the structure of the disk. Though we lack radial velocity measurements
and therefore do not have a precise mass value for the secondary, it
is possible to constrain its mass by using the period and duration of
the eclipse to put limits on the Hill radius, the point at which the
gravitational influence of the secondary balances that of the primary. 
For a circular orbit,

\begin{equation}
\label{eqn:hill}
r_H = a\left(\frac{\mu}{3}\right)^{\frac{1}{3}},
\end{equation}

\noindent where $a$ is the semi-major axis and $\mu$ is the mass ratio
  $\mu \equiv M_2/\left(M_1 + M_2\right)$. The outer radius of a gas disk
  extends out to $\xi r_H$, where estimates of $\xi$ are typically
  between $\xi \sim 0.3$ \citep{Ayliffe09} and $0.4$ \citep{Martin11}. 
  The semi-major axis can be calculated using the orbital period and 
  the mass of the primary, giving $a = 367R_\odot = 1.7$ AU. The outer 
  radius of the disk can be calculated from the ratio of the eclipse 
  duration to the period

\begin{equation}
\label{eqn:angles}
\frac{T_{\mathrm{ecl}}}{P} = \frac{\Theta}{2\pi},
\end{equation}

\noindent where $\Theta$ is the angular distance traversed by the 
secondary/disk system during the eclipse.

\begin{figure}[htp]
\epsscale{1.0}
\plotone{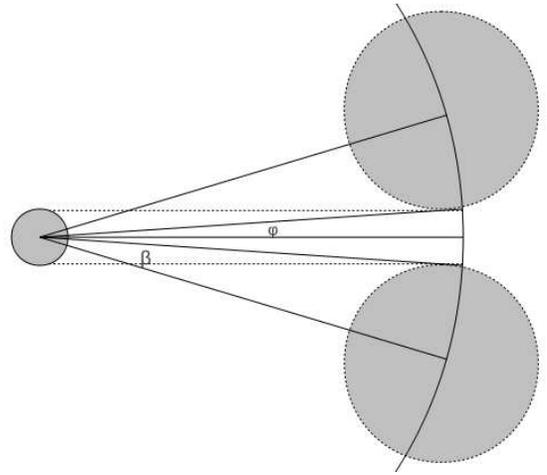}
\caption{Schematic of the total angular distance traversed by the
  secondary/disk system over the duration of the eclipse. Since only a
  small fraction ($\sim3\%$) of the period is taken up by the eclipse,
  the angles are small.
\label{fig:eclipse}}
\end{figure}

As seen in Figure \ref{fig:eclipse}, $\Theta = 2\varphi + 2\beta$. 
Since the eclipse-to-period ratio is small, we use the small angle 
approximation $\varphi \simeq$ sin($\varphi$) $\simeq R_1/a$ and 
$\beta \simeq$ sin($\beta$) $\simeq \xi r_H/a$, giving $\Theta = 
\left(2\xi r_H + 2R_1\right)/a$. Inserting this and Equation 
(\ref{eqn:hill}) into Equation (\ref{eqn:angles}), we obtain the 
cubic equation

\begin{equation}
\label{eqn:secmass}
\left(\xi^3 - 3\pi^3\frac{T^3}{P^3}\right)M_2 + 3\xi^2k^{\frac{1}{3}}M_2^{\frac{2}{3}} + 3\xi k^{\frac{2}{3}}M_2^{\frac{1}{3}} + k - 3\pi^3\frac{T^3}{P^3}M_1 = 0,
\end{equation}

\noindent where $k = 12\pi^2R_1^3 / GP^2$. Solving for $M_2$, we 
obtain bounds\footnote{The upper mass limit is not a hard limit, 
as a disk can potentially be smaller relative to the Hill radius 
of the secondary.} on the secondary mass of $M_2 = 0.23 M_\odot$ 
for $\xi = 0.4$ and $M_2 = 0.61 M_\odot$ for $\xi = 0.3$, assuming 
a circular orbit. In addition, the lack of a visible secondary
eclipse puts an upper limit of 2.1 $M_\odot$ on the mass of the 
secondary for this system's age and metallicity.

The best-fit result for a gas disk was an edge-on disk ($\theta_y =
0^{\circ}$) with an obliquity of $\theta_x = 0.9^{\circ}$, an orbital
inclination of $i = 89.7^{\circ}$, an inner radius of $R_{\rm in}$ = 
2.4 $R_\odot$, an outer radius of $R_{\rm out}$ = 35.5 $R_\odot$, a 
secondary mass of $M_2 = 0.32 M_\odot$, a radial density dropoff of 
$\alpha = 2.8$, and a disk mass of log$\left(M_{\rm disk}/M_2\right)$ 
= -7.8. The best fit for a gas and dust rich model produced $\chi^2 
/ \nu = 1322 / 112 = 11.8$. The best-fit parameters for both models 
are listed in Table \ref{tab:fits}.

\subsubsection{Hybrid Disk \label{sec:hybrid}}

While both models show general agreement with the overall eclipse 
trend, both have problems fitting the finer details. Therefore, we 
attempted to see whether the fit could be improved by implementing 
a hybrid model that combined the flat dust disk and flared gas disk. 
The best fit for the hybrid model had an obliquity of $\theta_x = 
1.1^{\circ}$, a tilt of $\theta_y = 3.0^{\circ}$, an orbital inclination 
of $i = 89.74^{\circ}$, an inner radius of $R_{\rm in}$ = 16.7 $R_\odot$, 
an outer radius of $R_{\rm out}$ = 41.8 $R_\odot$, a secondary mass of 
$M_2 = 1 M_\odot$, a radial density dropoff of $\alpha = 1$, a disk 
mass of log$\left(M_{\rm disk}/M_2\right) = -9$ dex, a debris disk 
optical depth of $\tau = 6.5$, and a debris disk density dropoff 
power law of $p = 1.5$, which gave $\chi^2 / \nu = 1419 / 110 = 12.9$.

Setting the dust disk (which produced the better fit out of the two 
previous models) as the null model and the hybrid disk as the alternative 
model, we then used a likelihood ratio test to determine whether the 
hybrid model produced a statistically significant improvement in the 
fit. The likelihood function

\begin{equation}
\label{eqn:likelihood}
L = \prod_{i=1}^n\left(\frac{1}{2\pi\sigma_i^2}\right)^{1/2}\mathrm{exp}\left(-\frac{1}{2}\left(\frac{y_i-y(x_i)}{\sigma_i}\right)^2\right)
\end{equation}

\noindent \citep{Bevington69} becomes

\begin{equation}
\label{eqn:Lfinal}
L = \left(\frac{1}{\pi\sigma^2}\right)^{n/2}\mathrm{exp}\left(-\frac{\chi^2}{2}\right)
\end{equation}

\noindent when the errors of each data point are assumed homoskedastic. 
In order to evaluate whether the hybrid model produced a statistically 
significant improvement in the fit, we made use of the Bayesian Information 
Criterion (BIC)

\begin{equation}
\label{eqn:BIC}
\mathrm{BIC} = -2\mathrm{ln}\left(L_{\mathrm{max}}\right) + k\mathrm{ln}\left(N\right) = \chi^2 + k\mathrm{ln}\left(N\right)
\end{equation}

\noindent \citep{Liddle07, Schwarz78}, where $k$ is the number of free
parameters, $N$ is the number of data points, and $L_{\rm max}$  is the 
maximum likelihood.  The dust and hybrid models were fit with 7 and 9 
free parameters, respectively, to 120 data points, hence the dust and 
hybrid models had 112 and 110 free parameters, respectively. The best-fit 
debris disk model produces $\chi^2 = 1073.8$; therefore the hybrid 
model will have produced a statistically significant improvement only 
if it has a BIC lower than that produced by the dust disk model, or

\begin{equation}
\chi_{\mathrm{hybrid}}^2 + 9\mathrm{ln}\left(120\right) < 1073.4 + 7\mathrm{ln}\left(120\right),
\end{equation}

\noindent which gives $\chi_{\rm hybrid}^2 / \nu < 1064.2 / 110$ for the 
hybrid disk model. Since the best-fit hybrid model produced $\chi^2 
/ \nu = 1073.8 / 110$, we conclude that the hybrid model does not 
produce a statistically significant improvement in the fit.

\section{Discussion}

OGLE 11893 appears to be near the end of its main sequence life, and 
it is unlikely to have a companion that has retained its primordial
protoplanetary disk.  This leaves the possibility that the companion
is surrounded by either a dusty debris disk or a second-generation
gas-rich disk.  The disk-eclipse system OGLE 11893 bears a strong
resemblance to EE Cep, the only other known system discovered to date
with a disk eclipsing a Be primary. \citet{Galan12} speculate that
the circumsecondary disk of EE Cep is fed by mass loss from the 
primary Be star, and given the similarity of the two systems, this 
is a possibility for OGLE 11893 as well.

The simplest flat debris disk model produced the best fit to the data 
(see Figure \ref{fig:model}), with a $\chi^2 / \nu$ value of $1073.8 / 
112 = 9.6$. Even the best-fit parameters reproduced the light curve 
poorly in the central region of the eclipse. Currently the primary is 
modeled as a perfect sphere even though it has the spectral signatures 
of a rapidly rotating Be star \citep{Dong14}. Modifying the model star 
to account for the elongation and gravitational darkening at the equator 
due to its rapid rotation would be an important step in improving the 
model.

OGLE 11893 and other eclipsing systems like it can be used to 
learn more about the composition and geometry of circumstellar 
disks through continued modeling efforts and photometric and 
spectroscopic observations. The discovery of this system in 
addition to J1407 and EE Cep provides evidence that eclipsing 
circumsecondary disk systems may not be as rare as previously 
thought, and that both archival searches and long-term photometric 
monitoring of large areas of the sky are likely to uncover more 
of them.

\begin{deluxetable*}{lcccl}[htb!]
\tabletypesize{\scriptsize}
\setlength{\tabcolsep}{0.03in}
\tablewidth{0pt}
\tablecaption{Best Eclipse Fits for OGLE LMC-ECL-11893}
\tablehead{
{(1)}               & {(2)}         & {(3)}      & {(4)}      & {(5)}\\
{Parameter}         & {Dust}        & {Gas}      & {Hybrid}   & {Units}\\
{}                  & {Disk}        & {Disk}     & {Disk}     & {}}
\startdata
$\theta_x$          &          1.5  &        0.9 &        1.1 & deg\\
$\theta_y$          &          5.9  &        0.0 &        3.0 & deg\\
$i$                 &        89.46  &      89.70 &      89.74 & deg\\
$R_{\rm in}$             &         26.1  &        2.8 &       16.7 & \rsun\\
$R_{\rm out}$            &         44.8  &       35.5 &       41.8 & \rsun\\
$\alpha$            &           --  &        2.8 &        1.0 & ...\\
$M_2$               &         0.12  &       0.32 &       1.00 & \msun \\
log($M_{\rm disk}/M_2$) &         -6.9  &       -7.8 &       -9.0 & dex\\
$\kappa$            &          0.1  &        0.1 &        0.1 & m$^2$\,kg$^{-1}$\\
$\tau_0$            &         7.9   &        --  &       6.5  & ...\\
$p$                 &         0.6   &        --  &       1.5  & ...\\
$\chi^2 / \nu$      & 1073.8 / 112  & 1322 / 112 & 1419 / 110 & ...\\
\enddata 
\tablecomments{Final best-fit parameters for the dust disk and 
gas disk models. $M_2$ and $M_{\rm disk}$ were only varied in the 
gas disk model. In the dust disk model, secondary mass was held 
fixed and $M_{\rm disk}$ was calculated using the assigned opacity 
and the best-fit values of the disk dimensions.}
\label{tab:fits}
\end{deluxetable*}

\acknowledgements

E.L.S., E.E.M., and F.M. acknowledge support from NSF award AST-1313029.
E.E.M. and M.J.P. acknowledge support from NSF AST-1008908.
E.E.M., E.L.S., F.M., M.J.P., and C.P.M.B. acknowledge support from the 
University of Rochester College of Arts and Sciences.
This research has made use of NASA ADS, Vizier, and SIMBAD, and data
from the OGLE survey.
This publication makes use of data products from the Two Micron All
Sky Survey, which is a joint project of the University of
Massachusetts and the Infrared Processing and Analysis
Center/California Institute of Technology, funded by the National
Aeronautics and Space Administration and the National Science
Foundation.
The authors thank Subo Dong, Jose Prieto, Szymon Kozlowski, and Matt
Kenworthy for discussions, and we thank the referee for a thoughtful
and timely report.




\end{document}